\documentclass[%
 notitlepage,
 reprint,
 superscriptaddress,
 amsmath,amssymb,
 aps,
]{revtex4-2}

\usepackage{xcolor}
\usepackage{lipsum}
\usepackage{graphicx}% Include figure files
\usepackage{dcolumn}% Align table columns on decimal point
\usepackage{bm}% bold math
\usepackage{hyperref}% add hypertext capabilities
\usepackage{epstopdf}

\newcommand{\bra}[1]{\langle #1 \vert}
\newcommand{\ket}[1]{\vert #1 \rangle}
\newcommand{\norma}[1]{\vert #1 \vert^{2}}
\usepackage{ulem}

\begin{document}

\preprint{APS/123-QED}

\title{Nonlinear flip-flop quantum walks through potential barriers}
%\title{Nonlinear bit-flip quantum walker in programmable photonic circuits}% Force line breaks with \\
%\thanks{A footnote to the article title}%

%\author{Ann Author}
% \altaffiliation[Also at ]{Physics Department, XYZ University.}%Lines break automatically or can be forced with \\
%\author{Second Author}%
% \email{Second.Author@institution.edu}
%\affiliation{%
% Authors' institution and/or address\\
% This line break forced with \textbackslash\textbackslash
%}%
\author{F. S. Passos}
\affiliation{
	Instituto Federal de Alagoas, 57020-600, Macei\' o, Alagoas, Brazil.
}%

\author{A. R. C. Buarque}
\affiliation{Departamento de F\'isica, Universidade Federal de Pernambuco,
	50670-901 Recife, PE, Brazil }
\affiliation{
	Secretaria de Estado da Educa{\c c}\~ ao de Alagoas, 57055-055, Macei\'o, Alagoas, Brazil
}

% It is always \today, today,
             %  but any date may be explicitly specified
%The limiting distribution of quantum walks on lattices is a fundamental problem, with widespread applications in quantum information and computation. 
\begin{abstract}
The dynamics of nonlinear flip-flop quantum walk with amplitude-dependent phase shifts with pertubing potential barrier is investigated. Through the adjustment between uniform local perturbations and a Kerrlike nonlinearity of the medium we find a rich set of dynamic profiles. We will show the existence of different Hadamard quantum walking regimes, including those with mobile soliton-like structures or self-trapped states. The latter is predominant for perturbations with amplitudes that tend to $\varphi\rightarrow \pi/2$. In this system, the qubit shows an unusual behavior as we increase the amplitudes of the potential barriers, and displays a monotonic decrease in the self-trapping $\varphi_c$ with respect to the nonlinear parameter. A chaotic-like regime becomes predominant for intermediate nonlinearity values. Furthermore, we show that by changing the quantum coins ($\theta$) a non-trivial dynamic arises, where coins close to Pauli-X drives the system to a regime with predominant soliton-like structures, while the chaotic behavior are restricted to a narrow region in the $\chi$-$\varphi$ plane. We believe that is possible to implement and observe the proprieties of this model in a integrated photonic system.
\end{abstract}
%\pacs{03.65.-w, 05.60.Gg, 03.67.Bg, 03.67.Mn}
%\keywords{Suggested keywords}%Use showkeys class option if keyword
                              %display desired
\maketitle

\section{Introduction}
Understanding the dynamic properties in the quantum walk protocol (QWs) is a relevant problem in modern physics and has attracted a lot of attention over the past few decades. Quantum walks are known to generally exhibit spread quadratically faster over time than their classical counterparts due to interference and superposition quantum processes \cite{Aharonov}. These characteristics makes them a powerful toolbox in the field quantum information processing and emerging quantum technologies \cite{venegas2012quantum,kempe2003quantum}. 

The two main segments of QWs are the continuous- and discrete-time quantum walks. Restricting to discrete-time quantum walks (DTQWs), these works as a cellular automaton \cite{meyer1996quantum, karski2009,costa2018quantum,huerta2020quantum,shikano2014discrete} chained to a set of rules able to simulate quantum mechanical systems following the devised quantum algorithm by Feynman \cite{feynman}. The dynamics of the DTQWs are based on the proprieties of the local quantum state instead of the regular quantum mechanics approach of the eigenstates and energy levels of the Hamiltonian. This scenario provided fertile soil for the emergence of models where DTQWs present themselves as a model of universal quantum computing \cite{childs2009universal,lovett2010universal,childs2013universal} with numerous branches of applications for graph isomorphism
 \cite{douglas2008classical,berry2011two,wang2015graph,qiang2021implementing}, quantum search \cite{PhysRevA.67.052307,foulger2015quantum,de2018controlled,meyer2013nonlinear,meyer2014quantum,portugal2013quantum}, and quantum simulations \cite{kitagawa2010exploring,chandrashekar2010relationship,mallick2017neutrino,mallick2019simulating}. Furthermore, DTQWs have been extensively demonstrated on several experimental platforms such as nuclear magnetic resonance systems \cite{ryan2005experimental}, photons in waveguides \cite{perets2008realization}, light using optical device \cite{PhysRevA.61.013410},
  trapped atoms \cite{karski2009quantum}, synthetic gauge fields in a three-dimensional lattice \cite{boada2017quantum} and superconducting qubits \cite{gong2021quantum}, to name a few.

Even though quantum mechanics is intrinsic linear, nonlinearity plays a major role in several structures and interplay the dynamics of information transport by enabling the wave packet, or another kind of information transport, to move as a solitonic wave, or even to be self-trapped in a specific site in the lattice. Systems with nonlinear iterations are capable of generating linear DTQWs as observed in the the Gross-Pitaevskii equation of interacting particles in a Bose-Einstein condensates \cite{chandrashekar2006implementing,alberti2017quantum,dadras2018quantum,groiseau2019spontaneous,xie2020topological} and in relativistic quantum mechanics through  the Dirac equation \cite{huerta2020quantum,chandrashekar2010relationship,lee2015quantum}.  Besides linear walks created in a nonlinear environment, nonlinear models of DTQWs have attracted great attention due to their rich dynamics \cite{Navarrete2007, ChangWooLee2015, Buarque2019, Mendonca2020, Buarque2021, Vakulchyk2019, maeda2018weak,PhysRevA.106.042202}. The seminal model was presented by Navarrete \textit{et al.} \cite{Navarrete2007} where proposed an optical Galton board to mimic the action of a Kerr-like nonlinearity by a phase-gain in the coin operator, giving rise to a map of very characteristic solitonic behaviors and chaos. The effect of trapped states was investigated in \cite{Buarque2019} by Buarque and Dias. The authors showed in detail that there is a dependency between self-trapping and the quantum gates that act in the DTQWs protocol. Photonic systems \cite{gao2019implementation,bisianov2019stability,ding2017cross}, implementations in quantum information \cite{guo2011simplified} and quantum computing \cite{lin2009quantum} provides a few examples of extensions of this approach.

The spread of nonlinear quantum walker along the chain has been also analyzed in the presence of noises and disorder \cite{Buarque2021, Vakulchyk2019,vakulchyk2018almost}. In \cite{vakulchyk2018almost}, the authors considered a DTQW with phase disorder and a nonlinear quantum coin and showed the existence of stationary and moving breathers modes with superexponential space tails. Understanding the dynamics of qubits in systems with the presence of competition between nonlinearity and disorder is fundamental for the development of quantum technologies. In another hand, the physical systems may be manufactured with potential barriers along the chain of dispersion of the wave packet and quantum proprieties allows part of its wave function tunnel throughout the barrier and keeping a portion behind. This model proposed by Wong \cite{WONG2016QIP,wong2016quantum} aimed to shed a light on the consequences of such potential barriers to the quantum formulation of Grover's algorithm.
	
With these ideas in mind, here we set out to track the dynamics of nonlinear flip-flop quantum walks in the presence of perturbing potential barriers. Flip-flop systems have attracted a lot of attention from the scientific community in recent times \cite{PhysRevA.105.062417,PhysRevA.99.042129,marevs2020quantum,PhysRevApplied.6.044001,PhysRevB.106.165302}. Flip-flop qubits have been presented as a promising model for new quantum computing design in several studies \cite{tosi2017silicon, rei2022parallel, ferraro2022universal}. We investigate the role played by the interchange and the balance between the nonlinear parameter and the potential barriers in the spread  of the wave packet of the qubit. We can find an absence of diffusion, \textit{i.e.} self-trapping, for the Hadamard coin, due to the action of barriers, controlled with the parameter $\varphi$, and a particular kind of nonlinear interaction, that preserves the shape and keeps a significant probability of finding the wave function on the initial site of the chain. The harmony of the parameters has a thin line, and the system, for the right set of values, undergoes a chaotic-like state, even for high $\chi$, showing that the nature of self-trapping is not claimed only by the operation of nonlinear parameter which has the propriety of lock the spread of the wave packet in discrete nonlinear Schrödinger equation systems, as in continuous time quantum walk.

\section{Discrete quantum walk protocol}
The quantum walks are defined on the combination of the degree of internal freedom (called coin space) with the real positions space, \textit{i.e}, $H=H_{P}\otimes H_{C}$. The internal states of the qubit is usually described by a 2-vector $\{\ket{\uparrow},\ket{\downarrow}\}\in H_{C}$, where they represent the components of the wave-function in the spin up/down, respectively.
The position Hilbert space is $H_{P}=\{|n\rangle\}_{n=1}^{N}$  for a lattice with $N$ sites.
The generic initial state of the particle localized around a position $n_{0}$ can be written as
$
\ket{\Psi(t=0)} = (a_{\uparrow}\ket{\uparrow} 
+ b_{\downarrow}\ket{\downarrow})\otimes\ket{n_{0}},
$
where $a_{\uparrow}$ and $b_{\downarrow}$ are complex amplitudes that satisfy normalization $P=\norma{a_{\uparrow}} + \norma{b_{\downarrow}} =1$. 

The evolution is governed by the discrete-time master equation
$\ket{\Psi(t)}=\hat{U}(t)\ket{\Psi(t-1))}$,
where $\hat{U}$ is the propagator operator; and is given by $\hat{U}=\sum_{n}\hat{S}\cdot(\hat{C}\otimes\mathbb{I}_{P})$. 
The quantum coin operator $\hat{C}$ acts only in the coin space $H_{C}$ is projected in each site $n$. We can define the coin as arbitrary $SU(2)$ given by
\begin{eqnarray}\label{coin_operator}
\hat{C}&&=\cos\theta\hat{Z}+\sin\theta\hat{X},
\end{eqnarray}
here, $\theta\in [0,2\pi]$ controls the chirality of the quantum coin and $\hat{Z},\hat{X}$ are Pauli operators. For $\theta=\pi/4$ we recovered the well-known Hadamard operator.

The flip-flop qubit's conditional shift operator, with unitary hopping probability, is usually defined by
\begin{eqnarray}\label{equation_shift}
\hat{S} =\ket{\uparrow}\bra{\downarrow}\otimes \ket{n+1}\bra{n} + \ket{\downarrow}\bra{\uparrow}\otimes\ket{n-1}\bra{n}.
\end{eqnarray}
In our model,  we modified the flip-flop conditional shift operator to bound the particle to the local states with amplitude $\beta$, and shift, left/right, with amplitude is $\alpha$, i.e, $\hat{S}_{p}\rightarrow \alpha\hat{S} + \beta(\hat{I}\otimes\ket{n}\bra{n})$, this means that the operator acting upon the single qubit has probability $|\alpha|^2$ to act just like the regular flip-flop operator and $|\beta|^2$ chance of keeping the qubit in its state without any change in it. One can write the operators action, upon a given spin state, as:
\begin{eqnarray}
\hat{S_{p}}\ket{\uparrow}\otimes\ket{n}=\alpha\ket{\downarrow}\otimes\ket{n+1} + \beta\ket{\uparrow}\otimes\ket{n},\nonumber \\
\hat{S_{p}}\ket{\downarrow}\otimes\ket{n}=\alpha\ket{\uparrow}\otimes\ket{n-1} + \beta\ket{\downarrow}\otimes\ket{n}.
\end{eqnarray}
Futhermore, $|\alpha|^{2} + |\beta|^{2}=1$ with $\alpha\beta^{*} + \beta\alpha^{*}=0$; to keep the unitary conditional shift operator. 
The walk will be perturbed during the shift procedure.
We parametize, $\alpha=\cos(\varphi)$ and $\beta=i\sin(\varphi)$. For $\varphi=0$, we retrieve the flip-flop shift operator \cite{WONG2016QIP}.

Moreover, the quantum walker acquires an intensity-dependent phase by each of the spinor component at each instant of time. 
The propagator operator, at a given time, is written in a general formulation as $\hat{U}^{t}=\sum_{n}\hat{S}_{p}(\hat{C}\otimes\mathbb{I}_{P})K^{t-1}$. The $ K^t$ operator induces nonlinear effects according to nonlinear self-phase modulation based Kerrlike effect and is defined as
\begin{eqnarray}\label{nonlinear_operator}
\hat{K}^{t}=\sum_{n}\sum_{s=\uparrow,\downarrow} e^{i2\pi\chi P_{n,s}(t)}\ket{n,s}\bra{s,n},
\end{eqnarray}
where $\chi$ describes the nonlinear interaction strength with the medium and $P_{n}(t)=\sum_{s=\uparrow,\downarrow}|\langle n,s\ket{\psi(t)}|^{2}$ of finding the walker in the position $n$ in time $t$, the linear case is easily recovered considering $\chi=0$. This operator was introduced in \cite{Navarrete2007}.
With the model description in hand, we are ready to perform a detailed investigation into the  dynamics of the nonlinear discrete quantum walks through potential barriers, represented in the non-symmetric action of the shift operator.

%#########################################
%%############ RESULTS ###################
%#########################################

\section{Results} 
\begin{figure}
    \centering
    \includegraphics[width=\linewidth]{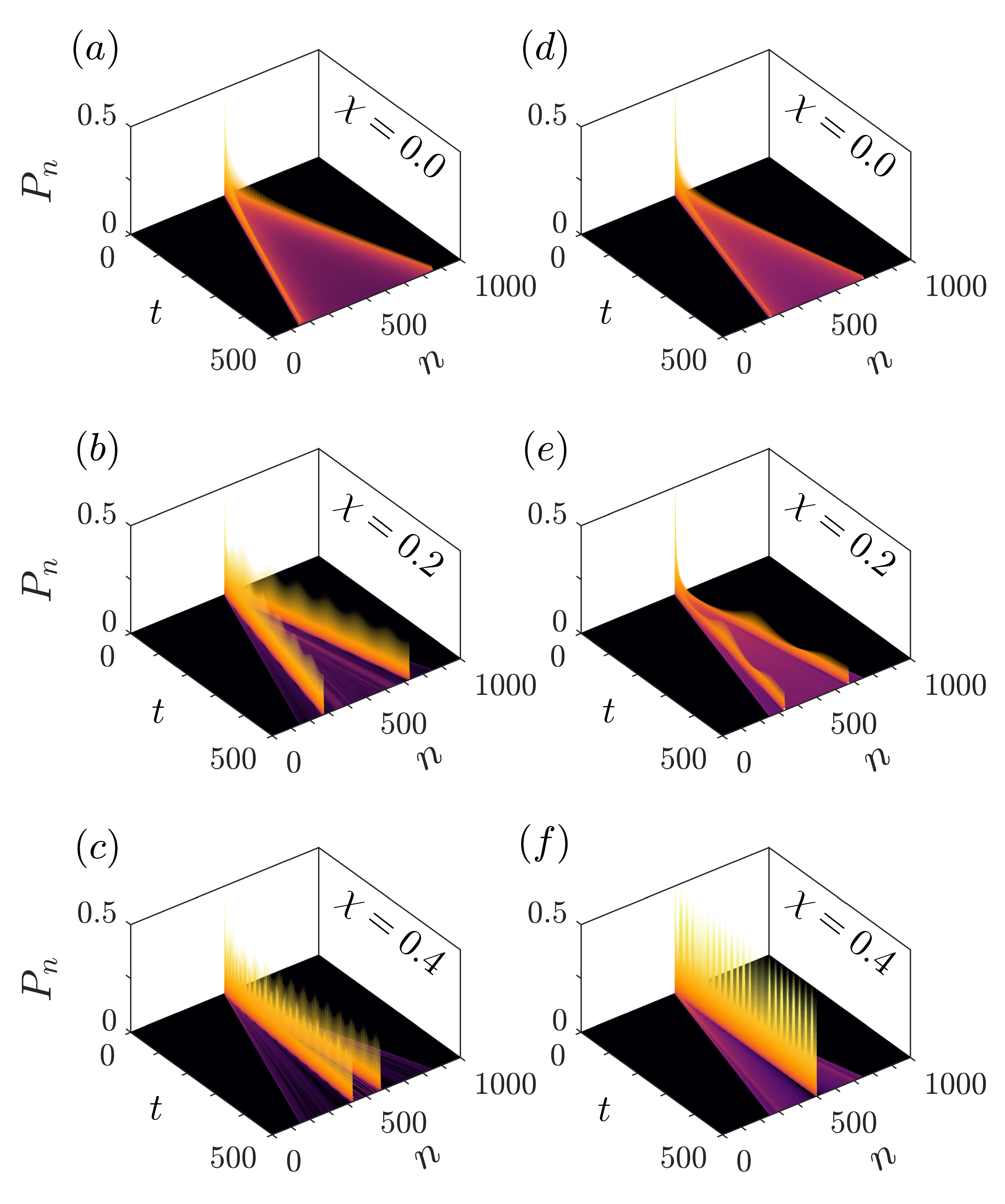}
    \caption{Snapshot of the space-time evolution of the probability density $P_{n}$ for some representative configurations between parameters $\chi$ and $\varphi$. The left column represents the case without potential barriers, $\varphi=0.0$. The right column displays a new dynamic when the quantum walker senses potential barriers with amplitudes $\varphi=\pi/4$.}
    \label{figure1}
\end{figure}

For all the results presented below, we consider as an initial condition a symmetric superposition between the states $\{\uparrow,\downarrow\}$ in the site $n_{0}=N/2$ of a $N$ sized chain, 
\begin{eqnarray}\label{initial_state}
\ket{\Psi(t=0)}=\frac{1}{\sqrt{2}}(\ket{\uparrow} + i\ket{\downarrow})\otimes\ket{n_{0}}.
\end{eqnarray}
We will keep the quantum coin fixed at $\theta=\pi/4$ (Hadamard walk) in order to investigate the competition between the nonlinearity and potential barriers.

We shall first track the evolution profile of the above state for Hadamard walk for some representative settings between nonlinearity and potential barriers. We show the time evolution of the probability density
 $P_{n}$ for lattice with $N=1000$ sites ruled by a Hadamard quantum coins in the absence of potential barriers $\varphi=0$ [left column, Fig. \ref{figure1} (a-c)] and in the presence of barriers with amplitudes $\varphi=\pi/4$ [right column, Fig. \ref{figure1} (d-f)]. 

In the absence of nonlinearity $(\chi=0.0)$, in both scenarios, the Hadamard coin induces a dynamic in which the qubit recover to the main features standard Hadamard walk, where the probability distribution shows two symmetrical peaks that ballistically spreads in opposite directions, see Fig. \ref{figure1} (a) and (d).
While in the potential-free walk ($\varphi=0.0$) the width between the peaks of the probability distribution appears in
$[-t/\sqrt{2}, t/\sqrt{2}]$,  when considering potential barriers the peaks occur at sites $[-\alpha t/\sqrt{2}, \alpha t/\sqrt{2}]$ \cite{WONG2016QIP}.  However, a keen eye notices that the parameter  $\varphi = \pi/4$ suggests that the  propagation velocity is slightly lower than the case where $\varphi=0.0$, due the presence of of the potential barriers that generates a new kind of propagation dynamics for the QW.
For $\chi>0$ the shape of the probability distribution is very different. In the presence of nonlinearity soliton-like structures emerge for any degree of nonlinearity strength \cite{Navarrete2007}. For low  degree of nonlinearity ($\chi=0.2$), we observe that the qubit exhibits a profile that is maintained (shape and amplitude) over time evolution. Now, the probability density of the qubit presents a concentrated behavior in two non-dispersive peaks while propagating along the line. Walking self-states have been reported in both DTQWs \cite{Navarrete2007, ChangWooLee2015, Buarque2019, Mendonca2020}.
On the other hand, when we consider the nonlinear interaction ($\chi=0.4$), a new phenomenology is observed, see Fig. \ref{figure1} (f). While in the absence of potential barriers ($\varphi =0.0$) the soliton-like structures are more evident, when the barriers have amplitudes $\varphi=\pi/4$ it is possible to see that, although the up and down states spread out in short time, for long time the two main amplitudes have a near zero velocity, engaging in a stationary state mode, with respect to peaks position, called self-trapped around initial site $n_{0}$, because it is induced by the nonlinear interaction of the qubit with the propagation medium, or mesh.
\begin{figure}
	\centering
	\includegraphics[width=\linewidth]{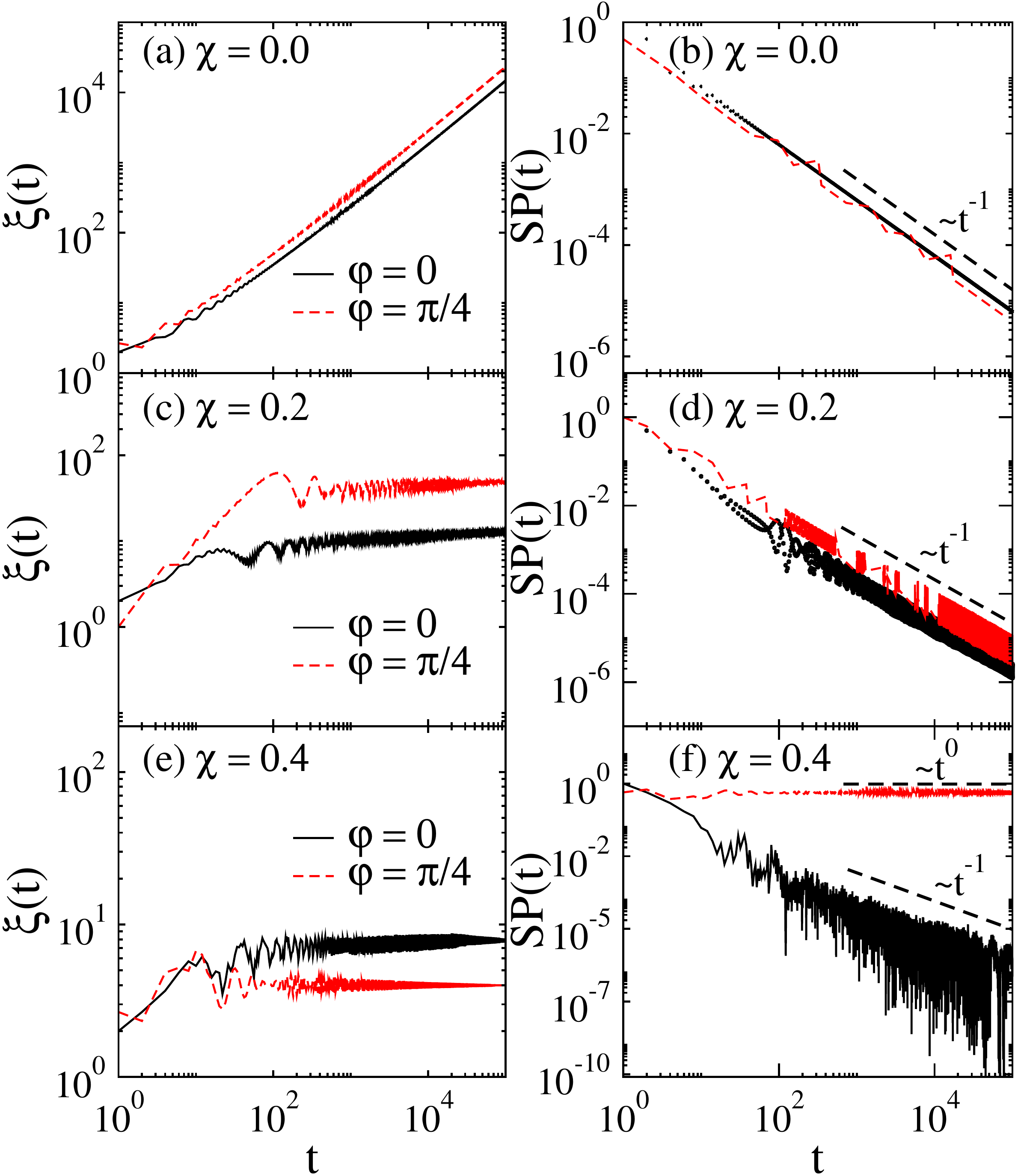}
	\caption{Time evolution of the participation function $[\xi(t)]$ and survival probability $[SP(t)]$ for three nonlinearity setting: (a)-(b) $\chi=0.0$, (c)-(d) $\chi=0.2$ and (e)-(f) $\chi=0.4$. In the absence of nonlinearity, the participation function displays an increasing behavior with time for the two values of $\varphi=0;\pi/4$ and a survival probability declines with a scaling behavior $SP\sim t^{-1}$. This scenario changes dramatically when we add nonlinearity. In (e) the participation function saturates in finite values, indicating the existence of localized states for the two configurations of potential barriers. The self-trapped quantum walk regime emerges for $\varphi=\pi/4$. The distinction between self-trapping walk and walking soliton-like structures is made through the fig. (f) where $SP\sim t^{-1}$ when $\varphi=0.0$, while the self-trapped states at the initial position $n_{0}$ presents $SP\sim t^{0}$.}
	\label{figure2}
\end{figure}

In contrast to the previous cases, when $\varphi=\pi/4$ the walker feels the potential barriers in the system, inducing a new dynamics. In this scenario, the probability density is greater around the initial site for the whole simulation time.
Previous studies did not report the self-trapping effect around the initial position for the Hadamard walk \cite{Navarrete2007,Buarque2019}. The system did not reveal this phenomenology for any configuration of nonlinearity \cite{Buarque2019}. In this context, the quantum walker presented solitonslike structures.

For a more quantitative description, we compute the time evolution of the participation function of the quantum walker in order to understand the effects caused by the nonlinearity in a walk through potential barriers.
The participation function is defined as
\begin{eqnarray}\label{participation_function}
\xi(t)=\frac{1}{\sum_{n}[P_{n}(t)]^{2}};
\end{eqnarray}
and gives the estimated number of chain positions over which wave-function visited at time $t$. For $[\xi(t)]$ increasing over time, we have a strong indication of delocalized states. Furthermore, when $\xi$ saturates in a finite value indicates a non spreading wave packet that could infer localized states.
Moreover, in order to analyze the dynamics of trapped states from a local perspective, we compute the evolution of the survival probability
\begin{eqnarray}\label{survival_probability}
SP(t) = P_{n}(t)\bigg\rvert_{n=n_{0}};
%SP(t)=\sum_{s=\uparrow,\downarrow}|\bra{s,n}\psi(t)\rangle|^{2}\bigg\rvert_{n=n_{0}};
\end{eqnarray}
this quantity describes an estimate probability of the quantum walker being found in the initial site $(n_{0})$ at time $t$. Thus, in the long-time regime the survival probability saturates at a finite value for a localized wave function, while $SP(t)\rightarrow 0$ means that the walker escapes from its initial location.
\begin{figure}
	\centering
	\includegraphics[width=\linewidth]{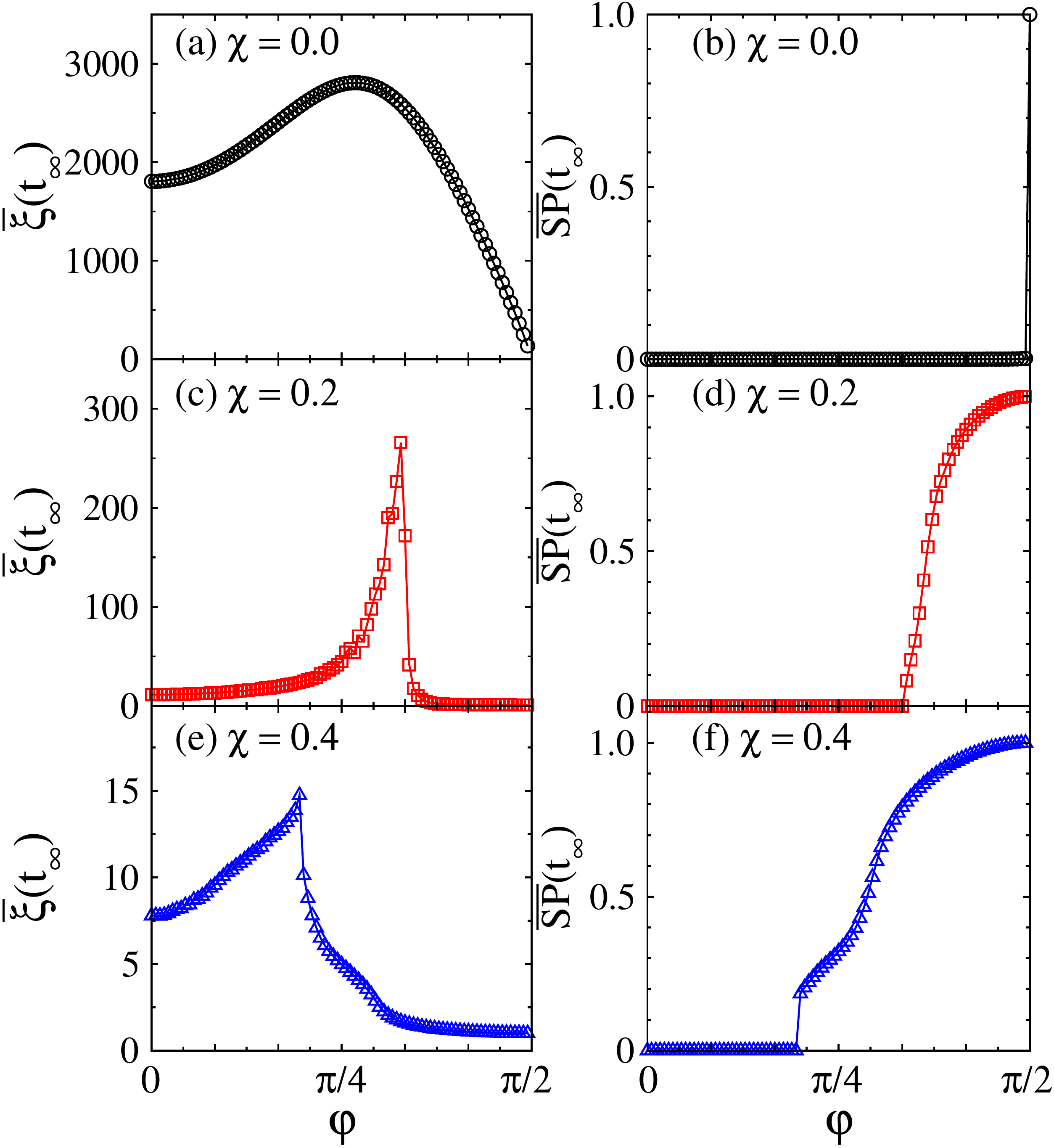}
	\caption{Long-time average of the participation function [$\overline{\xi}(t_{\infty})$] and survival probability [$\overline{SP}(t_{\infty})$] versus the potential barrier parameter $\varphi\in[0,\pi/2]$ for nonlinearity strength $\chi= 0.0$, $0.2$, $0.4$.
	In the absence of nonlinearity, the qubit wave packet spreads through the chain irrespective of the amplitudes of the barrier potentials, see (a-b). In the presence of nonlinearity ($\chi\neq 0$) the dynamics of the quantum walker is significantly changed.}
	\label{figure3}
\end{figure}

In figure \ref{figure2}, we consider the qubit wave packet initially at the center of the chain with $N=2\cdot 10^{6}$ sites and time steps $N/2$ in order to avoid edger effects. When the flip-flop quantum walk is purely linear, i.e, the walk free  of nonlinearity and in the absence potential barriers $(\chi=\varphi=0)$, respectively, in Fig. \ref{figure2}(a), we see that the states are spread across the chain's sites, since, $\xi(t)$ displays an increasing behavior in time for both cases ($\varphi=0$ and $\pi/4$). Furthermore, in Fig. \ref{figure2} (b) shows that the survival probability decreases monotonically with a scaling behavior $SP(t)\sim t^{-1}$, corroborating with  results previously in which the Hadamard walk $SP(t_{\infty})=2\pi/t$ \cite{xua2010}. When we add nonlinearity, the dynamics of the quantum walker are drastically changed.
Now, the probability density of the quantum walker display soliton-like structures for any degree of nonlinearity, i.e., permanent particlelike stationary state. In Fig. \ref{figure2} (c) shows the behavior of $\xi(t)$ for low degree of nonlinearity $\chi=0.2$. The participation function exhibits an approximately stationary behavior. On the other hand, the survival probability ($SP$) exhibit a qualitatively different dynamic behavior. The scaling law $SP(t)\sim t^{-1}$ suggesting mobile self-trapped states. For nonlinear interaction with intensity $\chi=0.4$, the participation function suggests the emergence of a dynamic with self-trapped states for both scenarios ($\varphi=0$ and $\pi/4$). However, when we look at the survival probability, we notice two dynamics for the quantum walker's self-trapped states. The results suggest that potential barriers ($\varphi=\pi/4$) induces the quantum walker self-trapping phenomenon in the initial site $n_{0}$. In this case, $SP$ displays a behavior that saturates at a finite value $SP(t)\sim t^{0}$, see Fig. \ref{figure2} (f). 

The above results suggest that the self-trapping regime is potential-barrier dependent. Now, we explore the asymptotic behavior of $\overline{\xi}(t_{\infty})$ and $\overline{SP}(t_{\infty})$ for Hadamard walk in a chain large enough to avoid border effects and averaged over the last 200 time steps. In Fig. \ref{figure3} we numerically analyze the dynamics for three representative values of nonlinearity strength: $\chi=0.0$, $0.2$ and $0.4$ for a range of potential barrier parameter contained in $\varphi\in[0, 2\pi]$.
For $\chi=0.0$, the early stage of potential barriers both $\overline{\xi}(t_{\infty})$ and $\overline{SP}(t_{\infty})$ suggest delocalized states. In this regime, the probability distribution of the quantum particle presents two peaks that depart ballistically from the initial position $n_{0}$.  The nonmonotonic increasing behavior the $\overline{\xi} (t_{\infty})$ arises due to local interactions for $\varphi>0$. When $\beta > \alpha$, local interactions begin to predominate under the quantum walker dynamics, inducing a localized behavior as $\varphi\rightarrow \pi/2$ and, therefore, $\overline{\xi} (t_{\infty})\rightarrow 1$. The survival probability is zero regardless of the value of $\varphi$, except $\varphi=\pi/2$ where dynamics are essentially localized, $\overline{SP}(t_{\infty})=1$, see Fig. \ref{figure3} (b).

In the presence of nonlinearity the profile of both physical quantities changes dramatically. 
In the regime with low amplitude potential barriers, the average participation function indicates dynamics are governed by self-trapped states. On the other hand, $\overline{\xi}(t_{\infty})$ displays growth in the vicinity of self-trapping transition. The average survival probability indicates that these are mobile self-trapped states, since $\overline{SP}(t_{\infty})\sim0$ for $\varphi<\varphi_{c}$, see Fig. \ref{figure3} (d).
Furthermore, in Fig. \ref{figure3} we note that critical amplitude above which the quantum walker's wave packet is trapped, exhibits a pronounced decrease as $\chi$ grows, that is, it suggests that $\varphi_{c}\rightarrow 0$.

\begin{figure}[t]
	\centering
	\includegraphics[width=\linewidth]{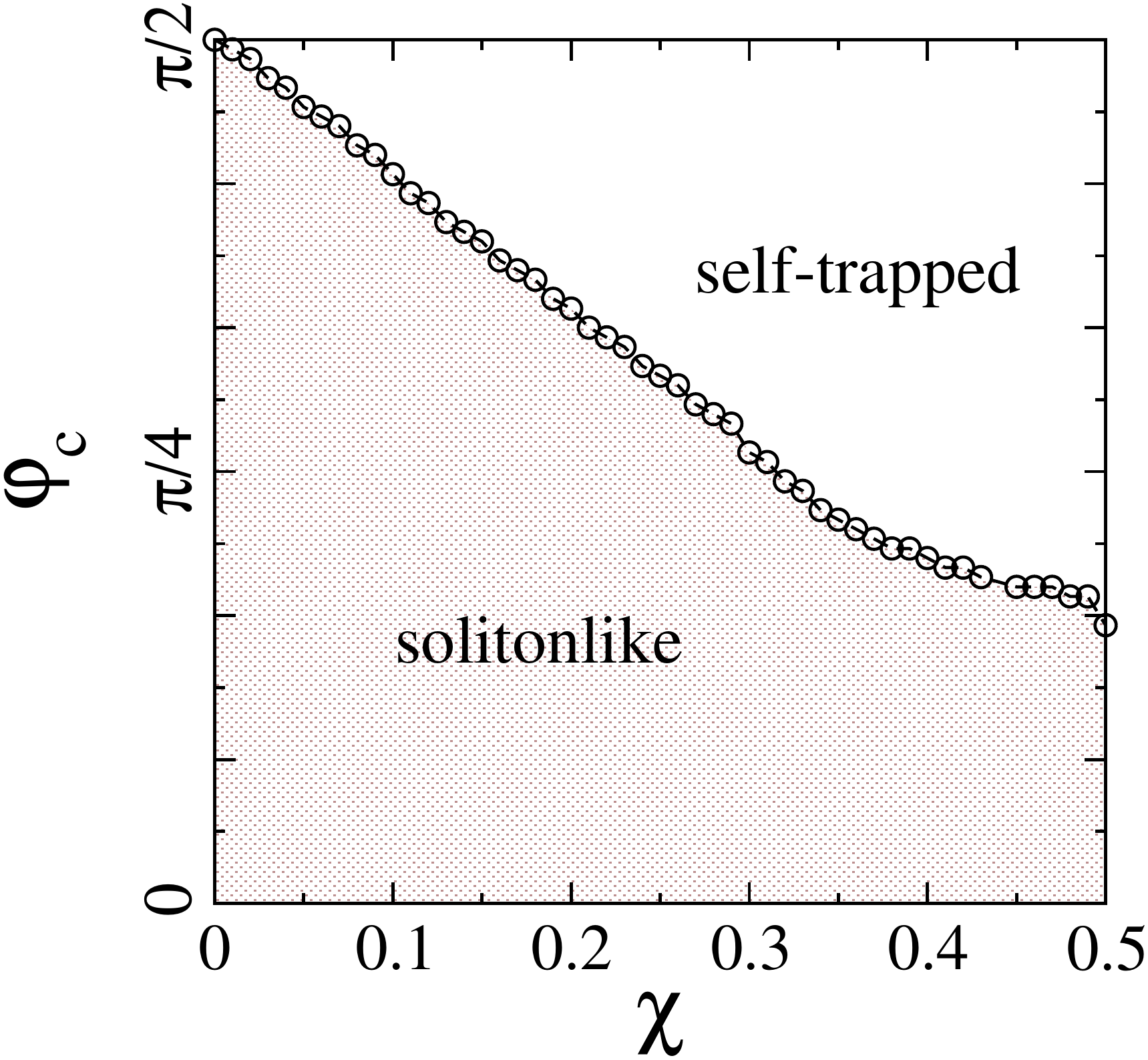}
	\caption{Phase diagram in the $\chi$ vs. $\varphi_{c}$ showing the transition between two regimes from the soliton-like to self-trapped. A monotonic decreasing of the critical value $\varphi_{c}$ as we increase the nonlinear parameter $\chi$.}
	\label{figure4}
\end{figure}

Now, to better understand this behavior of $\varphi_{c}$, we compute the time evolution for a long-time probability survival looking for configurations able to transition the quantum walker from the regime of mobile soliton-like structures to the self-trapping transition in initial position, $n_{0}$, restricted for $\chi\in[0,0.5]$, where the dynamics is correspondent the these specifications. 
In Fig. \ref{figure4}, the phase-diagram in $\chi$ versus $\phi_{c}$ showing the transition between these two well-defined regimes. A monotonic decreasing of local interaction $\varphi_{c}$ as we increase the nonlinear parameter $\chi$ is well observed.

\begin{figure}[t]
	\centering
	\includegraphics[width=\linewidth]{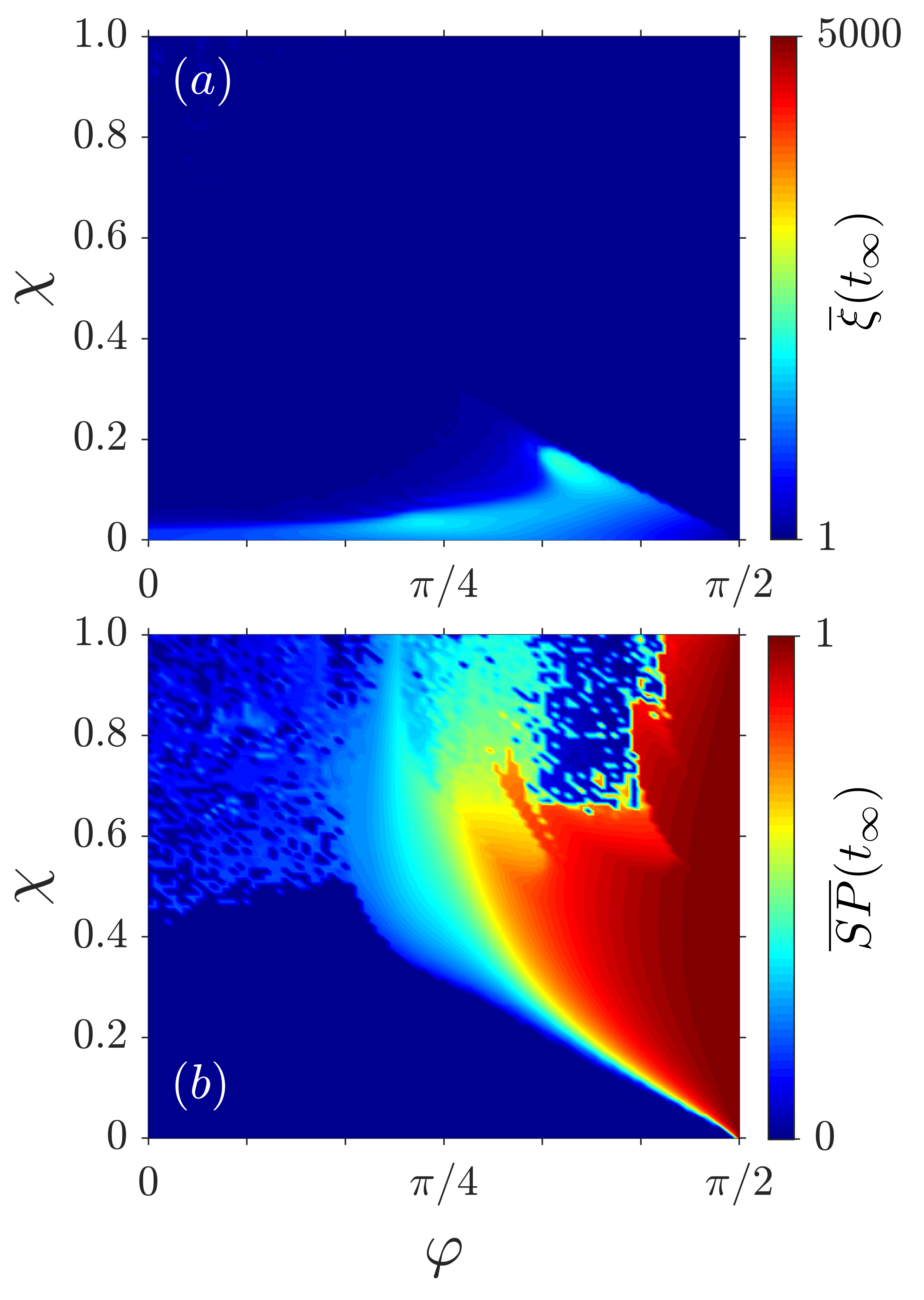}
	\caption{Plot of $\chi$ versus $\varphi$ for the long-time average of the participation function $\overline{\xi}(t_{\infty})$ (a) and survival probability $\overline{SP}(t_{\infty})$ (b). We note the absence of trapped structures for sufficiently small $\varphi$ values, even for a strong nonlinear parameter $\chi$. However, for $\chi>0.6$ a chaotic-like regime has been found. }
	\label{figure5}
\end{figure}

Finally, we extend our numerical experiments in order to offer a global view of the dynamical regimes. Fig. \ref{figure5} displays complete average behavior of the participation function [$\overline{\xi}(t_\infty)$] and probability survival [$\overline{SP}(t_\infty)$] in the $\chi$ versus $\varphi$ diagram, see Fig. \ref{figure5} (a) and (b), respectively.
When the Hadamard walk is purely linear ($\chi=0$) the dynamics is dispersive for all perturbing potential barrier with values in the range $0\leq\varphi<\pi/2$. However, the flip-flop qubit's  wave packet group velocity decreases as the extent to which $\varphi\rightarrow \pi/2$, since local disturbances become more intense. \cite{WONG2016QIP}. At the limit of low nonlinear interaction $\overline{\xi}(t_{\infty})$ indicates well-defined soliton-like structures that propagate ballistically in opposite directions, maintaining amplitude with small fluctuations.
\begin{figure}[t!]
%%	\centering
	\includegraphics[width=\linewidth]{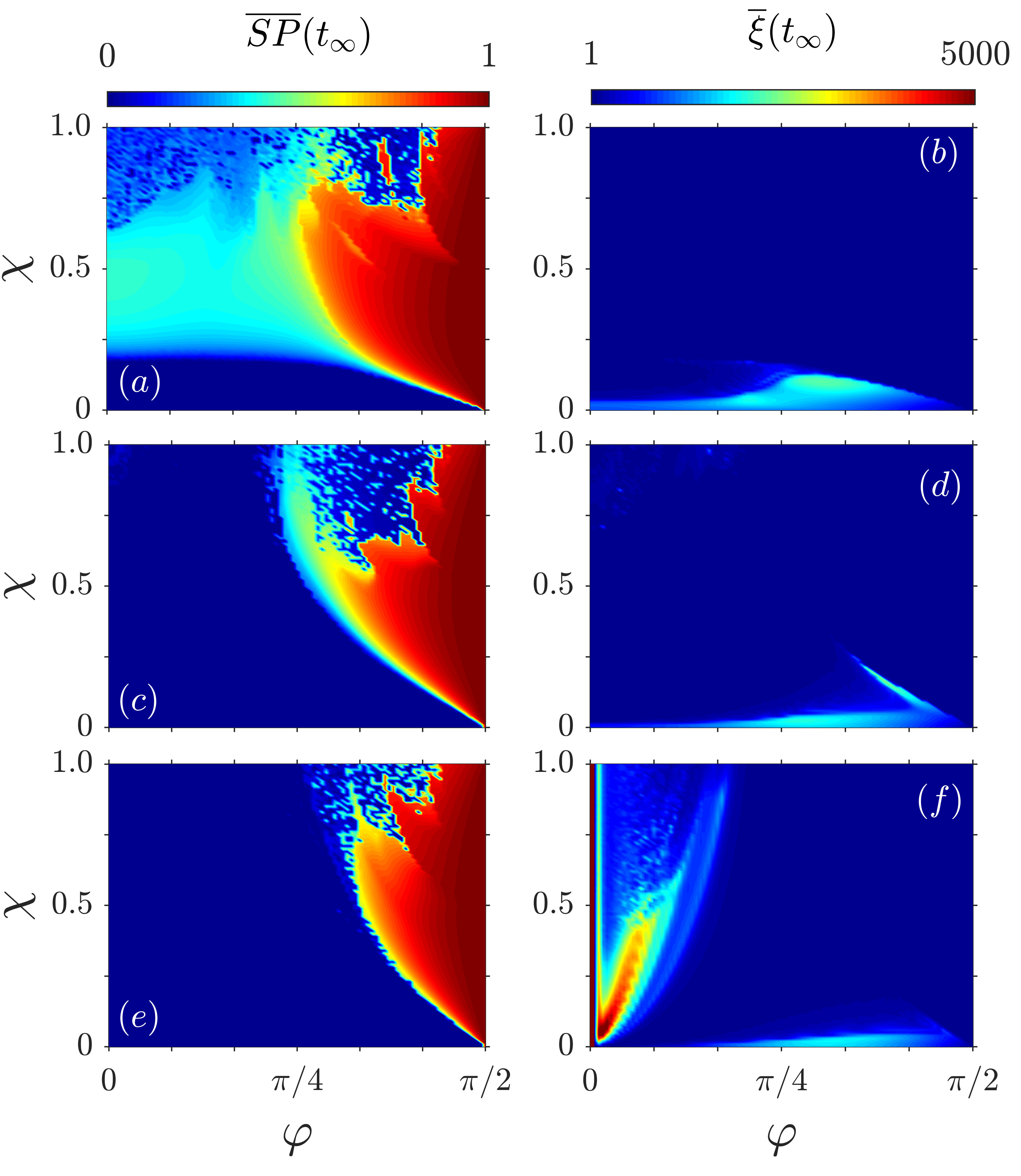}
	\caption{$\chi$ versus $\varphi$ diagram for the long-time average of the survival probability $\overline{SP}(t_{\infty})$ and also the long-time average participation function for three representative quantum coins (a)-(b) $\theta=\pi/6$, (c)-(d) $\theta=\pi/3$ and (e)-(f) $\theta=\pi/2$. The soliton-like phase suppresses the chaotic-like phase with the increase of $\theta$ and limits it to higher values of nonlinearity and in the low values an extended phase is observed in high $\theta$, as shown in (f). The self-trapping phase boundaries have little change in its length.}
	\label{figure6}
\end{figure}
For intermediate nonlinearity ($\chi>0.5$), the general dynamics of the quantum walker presents a chaotic aspect. The dynamics of the quantum walker is extremely sensitive to any fluctuations in the nonlinear parameter \cite{Navarrete2007}, as well as for any $\varphi$ values.
The self-trapping stationary quantum walk regime is even more concentrated when $\varphi\rightarrow \pi/2$. Previous results have shown that the self-trapping phenomenon in nonlinear DTQWs has an applied quantum gate-dependency \cite{Buarque2019}. However, this effect was absent in Hadamard walk. Now,  self-trapped Hadamard walk has a perturbing potential barrier dependency making the walker strongly trapped in the initial position.

For a deep analysis about the dynamics of nonlinear quantum walks over potential barriers, we shall now focus our attention on the analysis to other quantum coins, i.e, to different values of $\theta$. Figure \ref{figure6} displays the behavior of the long-time survival probability in the initial position [$\overline{SP}(t_{\infty})$] and the partitipation function in the same time length, for three representative values of coins: (a)-(b) $\theta=\pi/6$, (c)-(d) $\pi/3$ and (e)-(f) $\pi/2$, in the $\chi$ versus $\varphi$ diagrams. For quantum coin $\theta=\pi/6$ [Fig.\ref{figure6} (a)], the self-trapping regime is predominant for any potential barrier value. Here, we see two self-trapping profiles. In the region represented by light blue color the qubit probability distribution is alternating between self-trapped at the initial site ($n_{0}$) and the nearest-neighbor ($n_{0}\pm 1$), resulting in $\overline{SP}\approx 0.5$ and evidenced by a low value in participation, $\overline{\xi}\approx1$, which infers a nonspreading wave packet. In the predominated region by intense red color, still in Fig.\ref{figure6} (a), the quantum walker is totally self-trapped in the initial position ($n_{0}$), since $\overline{SP}(t_{\infty})\approx 1$. In the low nonlinearity limit ($\chi<0.2$), the qubit dynamics is governed by soliton-like structures that propagate in opposite directions from the initial position where $\overline{SP}\rightarrow 1$. By increasing $\theta$ the soliton-like structures begins to overcome the predominance and a more clear and distinguish frontier between these two behaviors is set upon the system, see Fig.\ref{figure6} (c)-(d). Another interesting behavior appears when $\theta$ results in the Pauli-X quantum coin, in the region of low nonlinearity and small barrier strength, low $\varphi$, the qubit exhibits a delocalized wavefunction, observed in Fig.\ref{figure6} (f). 
In this regime, the increase of $\overline{\xi}(t_{\infty})$ suggests the adequate control between $\chi$ and $\varphi$ can used as a mechanism capable of increasing the propagation of the qubit.	
The chaotic-like state, observed in Hadamard coin, in Fig.\ref{figure5} (b), begins to shorten and restricted to higher nonlinearity and $\pi/4<\varphi<\pi/2$. The control of chaotic regimes in DTQWs is of great interest to design new cryptosystems \cite{el2020quantum}.

\section{Experimental proposal setup}
\begin{figure}[t!]
	\centering
	\includegraphics[width=\linewidth]{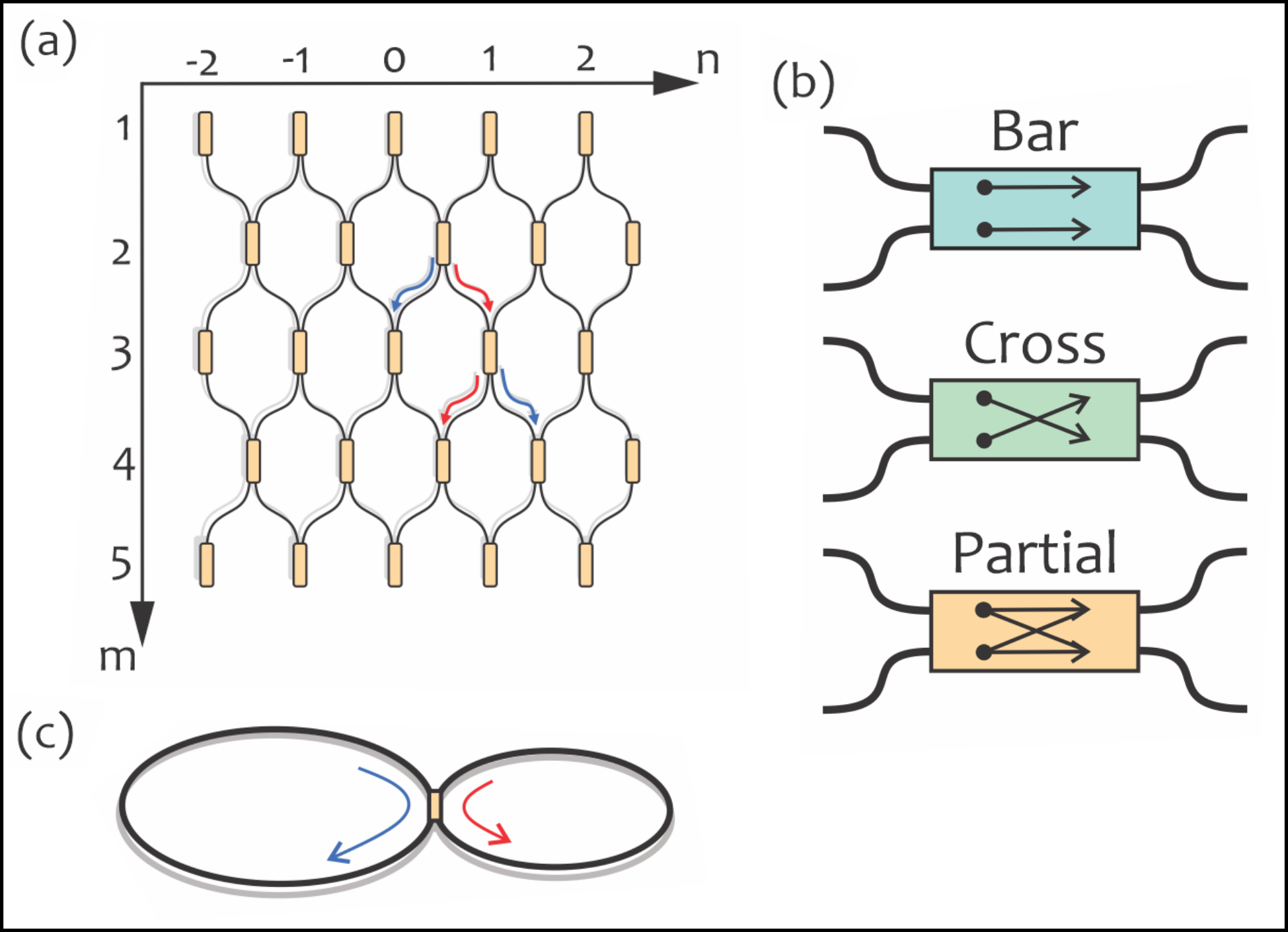}
	\caption{Universal 2x2 optical gate illustration. Two coupled-fiber loops of different lengths are connected by a fiber couple. In (c) the gate can be tuned between bar, cross and partial states}
	\label{figure7}
\end{figure}

A rich field in exploiting the dynamics and the theoretical background of quantum walks is integrated photonics\cite{grafe2016integrated,wang2020integrated,bogaerts2020programmable}, by manufacturing a meshgrid of coupled integrated waveguides, as illustrated in fig \ref{figure7} (a), it is possible to increase the range and propagation of quantum information protocol. 

The input field into the directional coupler, made of two waveguides, works by splitting the beam in two, acting as a regular beam splitter, and each one of the resulting fields has the desired amplitude. The connectivity of the optical mesh is done by inserting a programmable 2x2 optical gate with the possibility of being tuned as a logical knot for the states of bar, cross and partial\cite{bogaerts2020programmable}[fig \ref{figure7} (b)]. These gates can be integrated in order to act as algorithms and complex photonic circuits. The outline of applications to the designed circuits ranges from medical biosensors, to bank security and even traffic mobility with self driven cars \cite{bogaerts2020programmable}.

In order to mimic the action of a potential the meshgrid can be assembled with fiber loops having different lengths a short and long optical pathways [fig \ref{figure7} (c)]. The light evolution is described by the set of coupled equations\cite{Navarrete2007,regensburger2012parity}: 

\begin{eqnarray}
	u_{n}^{m+1}& =& \frac{1}{\sqrt{2}}(u_{n+1}^{m} e^{i\Gamma|u_{n+1}^{m}|^{2}} + iv_{n+1}^{m} e^{i\Gamma|v_{n+1}^{m}|^{2}}) \\
	v_{n}^{m+1}& =& \frac{1}{\sqrt{2}}(v_{n-1}^{m} e^{i\Gamma|v_{n-1}^{m}|^{2}} + iu_{n+1}^{m} e^{i\Gamma|u_{n-1}^{m}|^{2}})
\end{eqnarray}
Where the quantities $u^m_n$ and $v^m_n$ are, in the temporal domain, the amplitudes of pulses in the lattice position $n$ in the time step $m$, for left and right paths. The parameter $\Gamma$ is the nonlinear control parameter of gain and loss for the photon interaction with the different pathways.  

In this kind of arrangement, of pseudopotential action, it was already observed the presence of phenomenology of Bloch oscilations\cite{regensburger2012parity,wimmer2015observation}, Anderson Localization \cite{weidemann2021coexistence}, topological tunneling \cite{weidemann2020topological}, self-acceleration \cite{longhi2022non}, and invisible potential \cite{longhi2022invisible}.

\section{Final remarks}
In summary, we have studied the dynamic properties of nonlinear quantum walks with
potential barriers. We consider a Kerrlike nonlinearity that is associated with the acquisition of a phase, with intensity-dependence, applied to the quantum walks flip-flop model. Here, the flip-flop qubit has a probability $\alpha$ of tunneling to neighboring sites ($n\pm 1$) and $\beta$ staying at position $n$. We show that through the optimal adjustment between uniform local perturbations and nonlinearity of the medium, it is possible to reach the self-trapping state for the Hadamard quantum walk. We also have shown that the $\varphi_{c}$  threshold between soliton-like and self-trapped regimes displays a monotonic decreasing of perturbing potential barrier as we increase the nonlinear parameter $\chi$. For $\chi>0.5$, we find a chaotic-like dynamic where the behavior of the probability distribution is extremely sensitive to any change in parameters, thus, we have a non-trivial regime of characterization of $\varphi_{c}$.
The stationary self-trapped regime becomes predominant as $\varphi$ gets closer to $\pi/2$ where $\beta\rightarrow 1$ and the nature of the quantum walks becomes purely localized. Extending our studies to other quantum coin with $\theta\in[0,\pi/2]$, we show the emergence of incredibly complex dynamics. For $\theta=\pi/6$, the qubit wave packet displays a predominantly trapped dynamics for intermediate nonlinearity. Furthermore, adjusting $\theta$ for quantum coin values close to Pauli-X the soliton-like structures begins to overcome the predominance and a more clear and distinguish frontier between these two behaviors is set upon the system. The control between quantum coin parameter and potential barriers allows a reduction in the chaotic regime, leaving the region of soliton-like structures predominant.
Finally, considering that the application of flip-flop interactions has attracted a lot of attention from information science and quantum computing, we hope that our study can drive further investigations and applications of DTQWs in nonlinear photonic lattice. In the context of experimental platforms, we consider programmable photonic circuits systems as the most promising in the implementation of our results, considering that the core observations and descriptions could be observed in a relatively small setup, large enough to enable the QW to make a 100 steps walk, approximately. 

\section{Acknowledgments}

This work was partially supported by CNPq (Brazilian National Council for Scientific and Technological Development), via public notice 25/2021/PDJ. We also thank Wandearley Dias for the numerous suggestions, paving the way for the construction of this project.

\bibliography{references.bib}

\end{document}